\documentstyle[aps]{revtex}
\begin{document}
%
%

\def\bfk    {{\bf k}}
\def\kF     {\bfk_F}
\def\bcsgap {$\Delta_\bfk$ }
\def\GapA   {$\Delta_{\bfk,\bfk^\prime}$ }
\def\GapB   {$\Delta_{\bfk,{\bf 0}}$ }
\def\htc    {high-$T_c$ }
\def\lU     {large-$U$ }
\def\tJ     {$t-J$ }
\def\FS     {Fermi surface} 

%
%

\title{Correlation induced vanishing of the gap function on the Fermi
surface in \htc superconductors}

\author{G. Baskaran$^\ast$, D. G. Kanhere$^\dagger$,
        Mihir Arjunwadkar$^\dagger$ and Rahul Basu$^\ast$}

\address{${}^\ast$    The Institute of Mathematical Sciences,
                      Madras 600 113, India}
\address{${}^\dagger$ Department of Physics, University of Poona,
                      Pune 411 007, India}

\maketitle

\begin{abstract}

We present physical arguments as well as numerical evidence to show
that the Cooper pair (gap) function vanishes on the Fermi surface
in strongly correlated electron systems such as \tJ and \lU
Hubbard models in one and two dimensions.
Using exact diagonalization, we study a correlation function which is
a measure of the gap function even in a finite system,
and present numerical results for this correlation function on
small clusters of 8 (Hubbard) and 16 ($t-J$) sites.

\end{abstract}


%
%

The study of \htc cuprates
has brought to light a class of singlet superconductors
with properties rather different from the BCS type of
superconductor.
In this context, there have been intensive photoemission,
muon spin resonance and tunneling
studies of \htc superconductors in recent times,
as well as several theoretical suggestions.      \cite{nonBCS}
In particular, some of the recent models of \htc superconductors
have given rise to the possibility of gap functions
\begin{equation}
      \Delta_\bfk = \langle c_{\bfk \uparrow}^\dagger
                            c_{-\bfk \downarrow}^\dagger \rangle
\end{equation}
having unusual symmetry properties, e.g. \bcsgap which is {\em odd}
across the Fermi surface (FS), \cite{odd_k,pwa}
although the issue of the symmetry of the gap function still remains
unresolved.
In the context of the Resonating Valence Bond (RVB) theory,
which is based on strong electron correlations, one has always looked
for unique signatures in some of the physical properties.
The behaviour of the superconducting gap function in $\bfk$-space offers
one such signature.

Anderson \cite{pwa} has pointed out in several contexts that the gap
function vanishes on the FS and has the property
$sgn\Delta(\kF+\delta \bfk)=-sgn\Delta(\kF-\delta \bfk)$.
At simple mean-field level, this is true for the $t-J$ model only at
half filling (Mott insulator).
For the doped case, the analysis is difficult and Anderson did not
provide a convincing argument as to why the gap should vanish on the
FS away from half-filling.\cite{bza}
We must point out here that the possibility of odd behavior as a
function of $(|\bfk|-\kF)$ was anticipated by Cohen \cite{cohen}
many years ago, and recently many authors \cite{odd_k} have looked
at a more general type of singlet pairing that is odd in $\bfk$ and
$\omega$.

The aim of the present letter is to provide physical arguments
and numerical evidence for this claim.
We point out that the singular forward scattering between two
electrons with opposite spins on the FS, arising from strong
correlation, has an interesting consequence on the structure
of the gap function around the FS;
in particular, it has a tendency to vanish on the FS, independent
of the rotational symmetry in $\bfk$-space.
Our numerical studies on small clusters yield results which are
indeed consistent with these expectations.          \cite{ExpGap}
These results have been obtained through exact diagonalization on
small clusters in 1D (8-site Hubbard and 16-site \tJ chains) and
in 2D ($\sqrt{8} \times \sqrt{8}$ Hubbard and 4 $\times$ 4 \tJ
planes).
Within the finite-size limitation, we clearly see a change of sign
of the gap function along radial directions across the FS, indicating
the presence of a line of zeros close to the FS, and furthermore,
it has $s$-wave symmetry. {\em The key argument of this paper is that
the nature of charge fluctuations in $k$-space imposed by real space
repulsions is incompatible with the nature of charge fluctuations
demanded by a finite $\Delta_\bfk$ on the FS}.

%
%

Our starting point is the one-band \lU Hubbard model or the
\tJ model, which are believed to describe the low-energy physics
of \htc superconductors.
Let us begin by noting that in the \lU Hubbard model, the on-site
pairing amplitude
$\Delta_{ii}=\langle c_{i\uparrow}^\dagger c_{i\downarrow}^\dagger \rangle$
is diminished considerably on account of the restriction on double
occupancy. In the \tJ model, of course, double occupancy is projected
out. Thus
\begin{equation}
  \Delta_{ii}= \langle
                  c_{i\uparrow}^\dagger c_{i\downarrow}^\dagger
               \rangle = 0
\end{equation}
which reduces, in $k$-space, to
\begin{equation}
  \Delta_{ii} = \sum_k \langle
                          c_{\bfk\uparrow}^\dagger
                          c_{-\bfk\downarrow}^\dagger
                       \rangle
              \equiv \sum_\bfk \Delta_\bfk = 0,
  \label{constraint}
\end{equation}
where we use the fact that
$\langle
  c_{\bfk\uparrow}^\dagger c_{\bfk^\prime\downarrow}^\dagger
 \rangle = \Delta_\bfk \delta_{\bfk,-\bfk^\prime}$.
This, of course, is a global constraint on the gap function which can
be satisfied in several ways.
For example, in the RVB mean field theory at half filling, \cite{bza}
the gap function is constant and changes sign across the pseudo FS
defined by $cos(k_xa)+cos(k_ya)=0$ and satisfies the constraint
$\sum_\bfk \Delta_\bfk =0$.
Thus, in general, one or more lines of zeros such that
$\sum_\bfk\Delta_\bfk=0$ will occur.

In general, there is no reason to expect \bcsgap to vanish on the
Fermi surface.
In a simple $d$-wave superconductor in a 2D square lattice, for example,
the gap goes to zero along the lines $k_x=k_y$ and $k_x=-k_y$ and these
lines intersect the FS at four points.
But, is there any special advantage for the line of zeros to coincide
with the FS in 2D and in 1D?
We will now argue that the answer is in the affirmative.

Let us begin by examining the behaviour of the BCS wave function
close to the FS:
\begin{equation}
   \vert BCS \rangle = \prod_\bfk (u_\bfk + v_\bfk
                                  c_{\bfk \uparrow}^\dagger
                                  c_{-\bfk \downarrow}^\dagger)
                                  \vert 0 \rangle
\end{equation}
which may be rewritten as
\begin{equation}
   \vert BCS \rangle = \prod_{\bfk}{}^\prime (u_\bfk^2 +
                       \sqrt{2} u_\bfk v_\bfk b_{\bfk,-\bfk}^\dagger +
                       v_\bfk^2 b_{\bfk,-\bfk}^\dagger b_{\bfk,-\bfk}^\dagger)
                       \vert 0 \rangle,
   \label{bcs}
\end{equation}
where the product is now over only half of the $\bfk$-space
(e.g. $k_x > 0$ and all $k_y$ in 2D), with
\begin{equation}
   b_{\bfk,-\bfk}^\dagger = \frac{1}{\sqrt{2}} (
                     c_{\bfk\uparrow}^\dagger c_{-\bfk\downarrow}^\dagger
                   - c_{\bfk\downarrow}^\dagger c_{-\bfk\uparrow}^\dagger )
\end{equation}
as the singlet pair creation operator on points $\bfk$ and $-\bfk$ in
$\bfk$-space.
It is clear that $u_\bfk^2$ is the {\em probability amplitude}
of finding no singlet pair with momentum $(\bfk,-\bfk)$,
$u_\bfk v_\bfk$ is that of finding one singlet pair
(of charge $2e$), and $v_\bfk^2$ that of finding
two singlet pairs (of total charge $4e$).
The BCS state has identical phase relations for various configurations
of pair occupancy in $\bfk$-space.
That is, when the product in Eq. (\ref{bcs}) is expanded out,
the resulting sum has identical phase for all terms, each term
corresponding to different configurations of the $(\bfk,-\bfk)$
occupancy.
Superconductivity can thus be thought of as a coherent charge-$2e$
fluctuating state in $\bfk$-space.
Since $u_\bfk v_\bfk$ is non-zero only in a thin energy shell around the
FS, the coherent $2e$ charge fluctuation is concentrated around the
FS.
(It is interesting to note that this coherence in $\bfk$-space
results in phase coherence among the Cooper pairs in real space also).
Away from the shell, we either have a completely filled band (inside the
FS) or a completely empty band (outside the FS) and
hence no charge fluctuations.
Thus, $u_\bfk v_\bfk (\sim \Delta_\bfk)$ is a measure
of coherent charge fluctuations in $\bfk$-space.

We now argue that strong correlations in real space lead to a suppression
of such coherent charge fluctuations close to the FS.
Strongly correlated electrons in 1D and 2D, described by a
\lU Hubbard model, have certain unique features close to the FS.
It is well known that in the 1D Hubbard model there is singular forward
scattering between two electrons with opposite spins close to the FS.
This leads to a finite phase shift \cite{sfs} at the FS and the consequent
failure
of the Fermi liquid theory, resulting in the vanishing of the discontinuity
in $n_\bfk$ at the FS (Luttinger liquid behaviour).
It also implies an effective hard-core repulsive pseudopotential between
electrons with opposite spins close to the FS.
Thus no two electrons close to the FS, with opposite spins, can have the
same momentum, thereby making $\bfk$-points close to the FS essentially
singly occupied.
Coherent pair fluctuations are thus unlikely to develop on or very close to
the FS, but are not forbidden away from the FS.
It is therefore likely that the line of zeros of \bcsgap implied by the
global constraint (Eq. (\ref{constraint})) will coincide with the FS.

Generalizations of the above argument to two and higher dimensions is
straightforward provided the following is true: the projective
consraint of no double occupancy in real space should lead to singular
forward scattering and consequent failure of Fermi liquid theory. As
discussed earlier, the single occupancy constraint around the FS in
$\bfk$-space is likely to force the $(d-1)$ dimension nodal surface of
$\Delta_\bfk$ to coincide with the $(d-1)$ dimensional FS. In
particular, the 2-d case can be understood in the spirit of Anderson's
tomographic Luttinger liquid picture \cite{TLL} where we have a
collection of 1-d chains in $\bfk$-space.

%
%

The above discussion can be quantified by studying the behaviour of
\bcsgap numerically, for which purpose we perform exact diagonalization
calculations on small clusters of 8 and 16 sites.
However, it is not possible for us to calculate \bcsgap directly, since
we have chosen to work in a number-conserving basis.
We thus define a correlation function \GapA as
\begin{equation}
   \Delta_{\bfk,\bfk^\prime} = \langle
                                  b_{\bfk,-\bfk}
                                  b_{\bfk^\prime,-\bfk^\prime}^\dagger
                               \rangle,
\end{equation}
where the average $\langle \ldots \rangle$ is the ground-state
expectation value.
This quantity is clearly a measure of coherent pair fluctuations
between states $(\bfk,-\bfk)$ and $(\bfk^\prime,-\bfk^\prime)$ in the
ground state.

Before presenting the results, let us briefly consider the behaviour
of \GapA in the BCS case, for which it can be exactly evaluated, giving
us
\begin{equation}
   \Delta_{\bfk,\bfk^\prime} = u_{\bfk} v_{\bfk}
                               u_{\bfk^\prime}^\ast v_{\bfk^\prime}^\ast
                         \sim  \Delta_\bfk \Delta_{\bfk^\prime}^\ast.
    \label{factorise}
\end{equation}
Thus if \bcsgap vanishes on the FS and is an odd function across the FS,
it can be seen that \GapA $> 0$ when $\bfk,\bfk^\prime$ both lie either
completely inside or completely outside the FS, \GapA $=0$ when either
of $\bfk,\bfk^\prime$ lie on the FS, and \GapA $< 0$ when one is inside
and the other is outside.
In a strongly correlated system, we do not expect a complete factorization
as in Eq. (\ref{factorise}), but a form
\begin{equation}
   \Delta_{\bfk,\bfk^\prime} \sim \Delta_\bfk \Delta_{\bfk^\prime}^\ast
                                   + g_{\bfk,\bfk^\prime},
\end{equation}
where $g_{\bfk,\bfk^\prime}$ is expected to be small when $\bfk$ and
$\bfk^\prime$ are well separated.
We shall therefore infer the behaviour of \bcsgap from exact
results on \GapA by assuming a factorization of the form above.

%
%

We have evaluated \GapA using exact diagonalization on 8-site
Hubbard and 16-site \tJ chains in 1D and $\sqrt{8} \times \sqrt{8}$
Hubbard and $4 \times 4$ \tJ planes in 2D, for all $\bfk,\bfk^\prime$
within the Brillouin zone.
Assuming that \bcsgap changes sign across the FS and a factorization
of the form Eq.(\ref{factorise}) we expect to see a change in the
sign of \GapA
as $\bfk$ is varied across the FS for each value of $\bfk^\prime$.
We present results for a fixed value of $\bfk^\prime=0$ and
vary $\bfk$ along appropriate direction(s) across the FS.
The $\bfk^\prime \neq 0$ results are qualitatively similar.

The results for 1D are particularly striking.
Figure 1 depicts \GapB for an 8-site Hubbard chain with $U=10$ for
0, 2, 4 and 6 holes.
A clear sign change is observed across $\kF$.
To contrast this with the $U<0$ Hubbard model, we have also plotted
\GapB for $U=-10$ with 2 holes in the same figure.
As expected, \GapA is always positive and nearly flat around $\kF$.
This is consistent with on-site pairing for $\frac{\vert U\vert}{t}>>1$,
which implies $\langle c_{i\uparrow}^\dagger c_{j\downarrow}^\dagger
\rangle \sim \delta_{ij}$ making \bcsgap virtually $\bfk$-independent.
Figure 2 depicts \GapB for a 16-site \tJ chain with 0, 2 and 12 holes
for $J=0.32$, which again shows a clear change of sign across $\kF$.
Figure 3 is a plot of $k_s$ vs. $k_F$ for the 1-d chains, where $k_s$
is the node of $\Delta_\bfk$. An approximate linear relationship is
clearly evident, with the spread being due to the degeneracy of the
ground state.
In two dimensions, we present results
for the $4 \times 4$ cluster in Fig. 4 and 5.

Figure 4 shows \GapB for the $4 \times 4$ \tJ cluster ($J=0.32$), again
along $(0,0)-({\pi \over 2},{\pi \over 2})-(\pi,\pi)$, for 0, 2 and 12
holes.
As before, \GapB changes sign across $\kF$ for all the three cases.
We have observed a similar sign change along the direction
$(0,0)-({\pi \over 2},0)-(\pi,0)$.
Figure 5 is a surface plot of \GapB for all
$\bfk=(k_x,k_y), -\pi \leq k_x,k_y \leq \pi$.
In addition to the sign change, this also indicates the angular
symmetry of \bcsgap.
As $\bfk$ is moved along a closed path around the origin, we do not
see any sign change in \GapB,
which clearly indicates that the order parameter has $s$-wave symmetry.
One consequence of this is that $d$-symmetry need no longer be
invoked to enforce $\sum_k \Delta_k =0$. The system prefers to have
$s$-symmetry and the sign change across the FS satifies the global
constraint.

%
%

To summarise, we have investigated the behaviour of the gap function \bcsgap
across the Fermi surface in the \tJ and \lU Hubbard models in one and two
dimensions, at and away from half-filling.
We have presented physical arguments for the vanishing of \bcsgap on the FS.
Exact diagonalization results on 8- and 16-site clusters, both in 1D and 2D,
clearly indicate a change of sign of \bcsgap across $\kF$, and in addition,
the 2D results have $s$-wave symmetry. Our results, which essentially
gives a gapless $s$-wave superconductor, has consequences for several
experimental results in the cuprate superconductors below $T_c$ - in
particular, the temperature dependence of the London penetration depth,
NMR relaxation rate, zero bias conductivity in tunneling experiments,
specific heat etc.  Some of these consequences are being investigated.

%
%

\acknowledgements

We would like to thank V. N. Muthukumar for many useful and
illuminating discussions.
M.A. acknowledges the Council for Scientific and
Industrial Research for financial support, and the Center for
Development of Advanced Computing (C-DAC) for computing facility.
Partial financial assitance was provided under Project No.
SP/S2/M-47/89 by the Department of Science and Technology and DST
project SBR 32 of the National Superconductivity Programme.

%
%

%
%

{\bf Figure Captions}

\begin{enumerate}

\item
\GapB as a function of ${\bfk \over \pi}$ for an 8-site Hubbard chain
($U=10$) for 0, 2, 4 and 6 holes.
The corresponding $\kF = {\pi \over 2}, {\pi \over 4}, {\pi \over 4}$
and 0 respectively.
For comparison, we also plot \GapB for $U=-10$ for 2 holes.

\item
\GapB as a function of ${\bfk \over \pi}$ for an 16-site \tJ chain ($J=0.32$)
for 0, 2 and 12 holes.
The corresponding $\kF = {\pi \over 2}, {3 \pi \over 8}$ and ${\pi \over 8}$
respectively.

\item
The $k_s$ vs. $k_F$ curve for the 1-d chains; $k_s$ is the node of
$\Delta_\bfk$. The straight line is the $k_s=k_F$ plot as a ``guide to
the eye''.
\item
\GapB as a function of ${\bfk \over \pi}$ for a 2D $4 \times 4$ \tJ cluster
($J=0.32$) for 0, 2 and 12 holes.
$\bfk$ is varied along the direction $(x,x)$, with $x = 0, {\pi \over 2}, \pi$.
$\kF$ corresponds to $x={\pi \over 2}$ for 0 and 2 holes.
For 12 holes, $(0,0)$ is completely inside the FS and the other two points
completely outside.

\item
A surface plot of \GapB as a function of ${k_x \over \pi},{k_y \over \pi}$
for the $4 \times 4$ \tJ cluster ($J=0.32$) for 2 holes.
$\kF$ corresponds to the set of points $(\pm {\pi \over 2}, \pm {\pi \over 2}),
(\pm {\pi \over 2}, 0), (0, \pm {\pi \over 2})$.

\end{enumerate}


\begin{references}
\bibitem[\ddag]{email}
email: baskaran@imsc.ernet.in, kanhere@parcom.ernet.in,
rahul@imsc.ernet.in
\bibitem{nonBCS}
J. A. Martindale {\em et. al.}, Phys. Rev. {\bf B47}, 9155 (1993);
W. N. Hardy {\em et. al.}, Phys. Rev. Lett. {\bf 70} 3999 (1993);
T. E. Mason {\em et. al.}, to appear in Phys. Rev. Lett.; Z.-X. Shen
{\em et. al.}, Phys. Rev. Lett. {\bf 70}, 1553 (1993); N. E. Bickers,
D. J. Scalapino and S. R. White, Phys. Rev. Lett., {\bf 62}, 961
(1989); D. J. Scalapino {\em et. al .}, Phys. Rev. {\bf B35},6694
(1987); C. Gros, Phys. Rev. {\bf B38}, 931 (1988); E. Dagotto, NHMFL
preprint, August 1993 - to appear in Rev. Mod. Phys;
P. Monthoux, A. Balatsky and D. Pines, Phys. Rev. {\bf B46},
14803 (1992); D. A . Wollman {\em et. al.}, Phys. Rev. Lett. {\bf 71},
2134 (1993); L. Chen and A.-M. S. Tremblay,  C.R.P.S. preprint
CRPS-93-18 - to appear in Proceedings of Santa Fe Conference on
Spectroscopies in Novel Superconductors, 1993, Santa Fe, New Mexico;
P. Choudhury - private discussion.
\bibitem{odd_k}
F. Mila and E. Abrahams, Phys. Rev. Lett. {\bf 67}, 2379 (1991);
V. L. Berezinskii, Pis'ma Zh. Eksp. Teor. Fiz. {\bf
20}, 628 (1974) [JETP Lett. {\bf 20}, 287 (1974)].
E. Abrahams, A. V. Balatsky, J. R. Schrieffer and P. B. Allen,
Phys. Rev. B {\bf 47}, 513 (1993);
\bibitem{cohen}
M. L. Cohen, Phys. Rev. Lett. {\bf 12}, 664 (1964).

\bibitem{pwa}
P. W. Anderson,  {\em Proceedings of the Workshop on
Fermiology of High--$T_c$ Superconductivity}, Argonne, 1991;
P. W. Anderson, Science {\bf 235} 1196 (1987).

\bibitem{sfs}
P. W. Anderson, Phys. Rev. Lett. {\bf 64}, 1839 (1990);
P. W. Anderson in the ``Princeton RVB book'' on \htc
superconductors, Chapter V - unpublished.
See also F. D. M. Haldane, Phys. Rev. Lett. {\bf 47} 1840 (1981);
H. Shiba and M. Ogata, Prog. Theor. Phys. Suppl. {\bf 108}, 265 (1992).

\bibitem{TLL}
P. W. Anderson, Phys. Rev. Lett. {\bf 65}, 2306 (1990);
P. W. Anderson, {\em ibid.} {\bf 66}, 3226 (1991).

\bibitem{bza}
G. Baskaran, Z. Zou and P. W. Anderson, Solid State Commun. {\bf 63},
973 (1987).

\bibitem{ExpGap}
The gap \bcsgap for one layer is related to the actual, experimentally
measured gap $\Delta^{exp}_\bfk$ as
$\Delta^{exp}_\bfk \sim t_\perp^2 \Delta_\bfk$,
where $t_\perp(k)$ is the electron hopping matrix element between two
adjacent $CuO_2$ layers;see also S. Chakravarty, A. Sudbo, P. W.
Anderson and S. Strong, Science {\bf 261}, 337 (1993).

\end{references}
\end{document}